\documentstyle[dina4p,11pt]{article}
\def\b{\bibitem}
\def\v{\vec{v}}
\def\ol{\overline}
\def\S{{\cal S}}
\def\l{\lambda}
\def\i{\infty}
\def\x{|x|}
\def\M{{\cal M}}
\def\C{{\bf C}}
\def\R{{\bf R}}
\def\Eb{{\bar{\cal E}}}
\def\E{{\cal E}}
\def\zb{\bar{z}}
\def\o{\omega}
\def\c{\cite}
\def\r{\ref}
\def\rh{\rho}
\def\f{\frac}

\def\s{\sqrt}

\def\be{\begin{equation}}
\def\ee{\end{equation}}
\def\la{\label}
\def\i{\infty}
\def\xb{{\bar{\xi}}}

\def\L{{\cal L}}
\def\Lh{{\frac{1}{2} L}}
\def\ba{\begin{array}}
\def\re{{\rm Re}}
\def\im{{\rm Im}}
\def\ea{\end{array}}
\def\t{\tau}
\def\slr{{SL(2,{\bf R})}}

\begin{document}
\newtheorem{theorem}{Theorem}
\begin{center}
\LARGE{The Ernst equation on a Riemann surface}
\vskip2.0cm
\large{D. Korotkin\footnote{Supported by Alexander von Humboldt
      Foundation.}} and \large{H. Nicolai}
\vskip1.0cm
II. Institute for Theoretical Physics, Hamburg University, \\
Luruper Chaussee 149,  Hamburg 22761, Germany
\end{center}
\vskip2.5cm
{\bf ABSTRACT.} The Ernst equation is formulated on an arbitrary
Riemann surface. Analytically, the problem reduces to finding solutions
of the ordinary Ernst equation which are periodic along the
 symmetry axis. The family of (punctured) Riemann surfaces
admitting a non-trivial Ernst field constitutes a ``partially
discretized'' subspace of the usual moduli space. The method allows
us to construct new exact solutions of Einstein's equations in
vacuo with non-trivial topology, such that different ``universes'',
each of which may have several black holes on its symmetry axis,
are connected through necks bounded by cosmic strings. We show
how the extra topological degrees of freedom may lead to an
extension of the Geroch group and discuss possible applications
to string theory.
\vskip1.0cm
\section{Introduction}
General investigations of axisymmetric stationary solutions
of Einstein's equations in vacuo are commonly based on
the Ernst equation \c{Ernst, Hoens}
\be
(\E+\Eb)(\E_{z\zb}-\f{\rh_z\E_{\zb} +\rh_{\zb}\E_z}{2\rh})=
2\E_z\E_{\zb}
\la{E}\ee
which constitutes the main part of Einstein's equations. Here,
the complex variable $z=u+iv$ parametrizes the two (space-like)
coordinates, on which the complex Ernst potential $\E(z,\zb)$, and
hence the metric depend. The real function $\rho (z,\zb )$ (related
to the 33-component of the 4-metric) is harmonic, viz.
\be \rh_{z\zb}=0   \la{boxrho}    \ee
and can thus be represented as the imaginary part of some (locally)
holomorphic function $\xi (z)$, i.e.
\be
\rho (z,\zb) = {\rm Im} \, \xi (z)      \la{rho}
\ee
Defining
\be
x(z,\zb) = {\rm Re}\, \xi (z)         \la{x}
\ee
we can rewrite (\r{E}) as follows:
\[ (\E + \Eb)(\E_{xx} +\f{1}{\rho}\E_{\rho} +\E_{\rho\rho})=
2(\E_x ^2 +\E_{\rho}^2) \]
or, in complex notation,
\be
   (\E + \Eb)(\E_{\xi\xb} - \f{\E_{\xi} -\E_{\xb}}{2(\xi -\xb)} ) =
2\E_{\xi}\E_{\xb}
\la{E1}\ee
At this point, one usually switches
from the coordinates $(z,\zb )$ to new coordinates $(x,\rho)$
(so-called Weyl canonical coordinates) by means of the conformal
reparametrization $z \rightarrow \xi (z)$. The full metric
of the Einstein manifold $\cal M$ corresponding to a given solution
of (\ref{E}) can then be represented in the form
\be
ds^2 =dl^2 +dL^2
\la{ds}\ee
where
\be
dl^2 =f^{-1} e^{2k} d\xi d\xb
\la{dl}\ee
\be
dL^2 =-f(dt + Ad\phi)^2 +f^{-1}\rho^2 d\phi^2
\la{dL}\ee
The variables $t$ and $\phi$ are the time and angular coordinates,
respectively, so that there are two mutually commuting Killing
vectors $\partial _t$ and $\partial _{\phi}$, one time-like and
one space-like. The functions
$f$, $A$ and $k$ depending on $(\xi, \xb)$ are determined from
\be
f={\rm Re}\E \;\;\;\;\;\;
A_{\xi}=2\rho\f{(\E -\Eb)_{\xi}}{(\E+\Eb)^2}\;\;\;\;\;\;
k_{\xi}= 2 i \rho\f{\E_{\xi}\Eb_{\xi}}{(\E+\Eb)^2}
\la{co}\ee
Observe that the first order differential equations can be
consistently solved because the compatibility with the complex
conjugate equations is guaranteed by (\r{E1}).

Although the change of variables from $(z,\zb)$ to $(x,\rho)$ is
purely local, a simple global topology has been assumed in all previous
studies of axisymmetric stationary solutions of Einstein's
equations \c{Hoens}. Namely, one takes the complex
coordinate $x+i\rho$ to parametrize the upper half plane,
such that $x \in {\bf R}$ parametrizes the symmetry axis and
$\rho \geq 0$ is a radial variable. The class of exact solutions
obtained in this way includes the well known
Schwarzschild and Kerr solutions, which provide the mathematical
background for black hole physics, as well as a host of
other solutions with unphysical features.
In this paper, we want to exploit the link between
(\ref{E}) and (\ref{E1}) at the {\it global} level by assuming
$(z,\zb)$ to be local coordinates on an arbitrary Riemann
surface rather than (part of) the Riemann sphere. The function
$\xi (z)$ is then no longer globally single-valued.
An important element in our construction
is the Kaluza-Klein type interpretation of the metric (\ref{ds})
as a model of matter-coupled gravity in two dimensions.
Then the part $dl^2$ of the metric
(\r{ds}) is interpreted as the metric of a two-dimensional
world-sheet with local coordinates $(z,\zb)$, while the Ernst potential
$\E$ is regarded as a complex ``matter field'' living on this
world-sheet together with the ``dilaton'' field $\rho$ and the
Liouville degree of freedom contained in $k(z,\zb)$.
On the basis of this interpretation, we will
construct new exact solutions of Einstein's equations topologically
equivalent to the product of the time axis (parametrized by $t$),
the circle $S^1$ (parametrized by $\phi$) and
(part of) some Riemann surface $\L$.
Even for the trivial solution $\E =1$, we obtain topologically
non-trivial flat manifolds with an arbitrary number of asymptotically
flat regions. To be sure, such manifolds will have singularities
of some kind. As it turns out, the singularities of the four-manifold
correspond to $\delta$-function curvature singularities
on the world-sheet. In the 4D interpretation, these can be viewed
as ringlike cosmic strings with negative tension (i.e. instead of a
deficit angle in the neighborhood of the string, there is now an
excess angle).
We believe that the solutions presented here may have some physical
interest since they provide novel examples of wormhole-type solutions
for Einstein's equations without matter.

To define the Ernst potential $\E$ on the Riemann surface $\L$ and
to exploit the global freedom left in the choice of $\xi(z)$, we
utilize a description of $\L$ introduced by Mandelstam \c{Man}
and further elaborated in \c{GW} for the computation of
multiple string scattering in the light-cone gauge. To this aim,
we choose a meromorphic abelian differential $d\xi(P)$ on $\L$
(points on $\L$ are labeled by $P,Q,...$) and define
\be
\xi (P) =\int^{P}_{P_0} d\xi      \la{xiP}
\ee
where the base point
$P_0 \in \L$ must not coincide with any singular point of
$d\xi (P)$, but otherwise can be chosen arbitrarily.
The real and imaginary parts
of $\xi(P)$ will be designated by $x(P)$ and $\rho (P)$ as in
(\ref{rho}) and (\ref{x}), respectively. The differential $d\xi (P)$
defines a flat metric on $\L$ through
\be
ds_0^2= d\xi \otimes d\xb
\la{ds0}\ee
Obviously, $ds_0^2$ is degenerate at the zeros and singular at the
poles of $d\xi(P)$. Moreover, the function $\xi(P)$ is in general not
globally defined on $\L$ since the corresponding abelian
integrals generically have non-zero cyclic periods. If we want to
define the Ernst potential $\E = \E\big( x(P),\rho (P)\big)$
on $\L$ by making use of the local equivalence of
(\ref{E}) and (\ref{E1}), we must impose additional conditions on $\L$
and the Ernst potential in order to render $\E(x,\rho)$ globally
single-valued. We will need two requirements.
\begin{enumerate}
\item While the cyclic periods of the differential $d\xi (P)$ do not
vanish in general, we can always arrange them to be {\it real} by the
addition of suitable holomorphic differentials (this defines
$d\xi$ uniquely). Consequently,
the function $\rho (P)={\rm Im} \, \xi (P) $ is globally defined
on $\L$; it plays the role of a global (light-cone) time in the
Mandelstam description of string scattering.
The single-valuedness of $\rho$ is also indispensable for
single-valuedness of the corresponding metric in four dimensions.
\item If the function $x(P) = {\rm Re} \, \xi (P)$ is not
globally defined, the Ernst potential must be periodic in its first
argument with a certain period $L$, i.e.
$\E (x,\rho)= \E (x+L,\rho )$. In addition we must assume the (real)
cyclic periods $C_j$ of the integral $\xi(P)$ to be ``quantized'',
i.e. $C_j = n_j L$ where $n_j\in {\bf Z}$ and $L\in {\bf R}$.
Under these conditions $\E$ extends to a globally single-valued
function on the surface $\L$.
\end{enumerate}

In the special case when all
cyclic periods of $d\xi (P)$ vanish, we can drop the second
requirement since then $d\xi(P) =d h(P)$ for some meromorphic
function $h(P)$ on $\L$. Periodicity for $\E(x,\rho)$ is not required
in this case, and for any solution $\E(\xi,\xb)$ of
(\r{E1}) we get a non-trivial and globally defined Ernst potential
$\E(h(z),\bar{h}(z))$ on $\L$. The topological degrees of
freedom can be analyzed by switching off the ``matter excitations'',
i.e. by putting $\E =1$; they are described by the ordinary
moduli space of the associated (punctured) Riemann surface\footnote{The
full ``moduli space of solutions'' corresponding to the gauge
equivalence classes of solutions in the presence of matter
is, of course, infinite-dimensional.}. If we now switch
on some non-trivial solution of (\r{E1}) with period $L$, the
quantization condition above will lead to a (partial) discretization
of moduli space since all cyclic periods
of the integral $d\xi$ are proportional to $L$ with integer
coefficients. Thus the Riemann surfaces amenable to our construction
constitute a ``partially discretized'' subspace of
ordinary moduli space. On the worldsheet $\L$,
we can get additional branch points
by choosing $\E( \xi, \xb)$ and/or
$f^{-1} e^{2k}$ with extra branch points in the $\xi$-plane.

The existence of topologically non-trivial solutions has
important implications for the symmetry structure of the theory.
As has been  known for a long time, the Ernst equation belongs
to the class of integrable equations. The first indication
of this basic property appeared in \c{G} where
it was shown that the Ernst equation admits an infinite-dimensional
symmetry group (Geroch group) acting on the space of axisymmetric
stationary solutions. The integrability of Einstein's equations
in this reduction was subsequently demonstrated in
\c{BZ} and \c{M}, where the associated zero curvature
representation (or Lax pair) was found. The
application of the methods of soliton theory by many authors allowed
the construction of multi-soliton solutions \c{BZ} and a study
of the associated Riemann-Hilbert problem \c{HE,Alex}. Furthermore,
the related B\"{a}cklund transformations were found in \c{N},
and a link between the Ernst equation and the deformation of
of hyperelliptic algebraic curves was established in \c{KM}.
The Kerr solution was interpreted in this framework as a special case
of the two-soliton metric. It should be kept in mind, however, that
most of the new solutions generated by this method are plagued by
unphysical features such as naked singularities and violations of
causality \c{Ves,NK,Kor} (this statement applies to both
stationary axisymmetric and colliding plane wave solutions;
see e.g. \c{Grif,Nic1} for reviews of the latter\footnote{The
colliding plane wave solutions can be formally obtained from the
axisymmetric stationary ones by a Wick-rotation (although their
physical interpretation is, of course, entirely different). In terms
of the notation adopted in \c{Nic1}, this amounts
to the replacement of $z,\zb,\xi (z)$ and $\xb $ by
by $x^+,x^-,\rho_+ (x^+)$ and $\rho_- (x^-)$, respectively.}).
The understanding of the underlying group theoretical structure
evolved from the early discovery of a ``hidden'' $\slr$ symmetry
in Einstein's equations to detailed studies of the
infinite solution generating symmetries of the axisymmetric
reduction \c{G2}. It advanced considerably with the advent of
(Kaluza Klein) supergravity and the discovery of hidden symmetries
in these theories \c{CJ}, which made it abundantly clear that
Einstein's theory is just a special example in a more general
class of (possibly supersymmetric) $G/H$ coset space sigma models
coupled to gravity. This led to the realization that the Geroch
group is nothing but (a non-linear realization of) the
loop group $\widehat {\slr}$ with a central extension acting as
a scaling operator on the conformal factor and that
the emergence of affine Kac Moody algebras in the reduction
to two dimensions is a general phenomenon \c{Julia, BM, Nic1};
in fact, even the $G/H$ coset structure of the higher-dimensional
theories has an infinite-dimensional counterpart.
On the basis of earlier conjectures that these affine Kac Moody
algebras admit hyperbolic extensions in the reduction to one
dimension \c{Julia2}, it was suggested in \c{Nic3} that there
should exist new symmetries associated with the topological
world-sheet degrees of freedom. Indeed, in this paper
we will exhibit a Virasoro-Witt algebra corresponding to the
variations of the conformal structure of $\L$ within the discretized
moduli space mentioned above. The full symmetry algebra is the
product of the Kac Moody algebra associated with the
Geroch group and this Virasoro-Witt algebra (this algebra has been
identified previously for topologically trivial world-sheets \c{M2},
but there the extra symmetries do not generate
new solutions beyond those generated by the Geroch group).
In addition, the Geroch group itself may also
change the topology of the worldsheet. This happens
whenever the function $\E(\xi,\xb)$ has a
singularity in the $\xi$-plane, where the conformal factor
$f^{-1}e^{2k}$ has a branch point; the worldsheet is then
a covering of $\L$ with these branch points.

This paper is organized as follows. In Section 2 we explain the
construction of flat metrics on an arbitrary Riemann surface and the
nature of their singularities. In Section 3 we apply these results
to the construction of topologically non-trivial flat four-dimensional
manifolds with cosmic string singularities. Periodic solutions
of the Ernst equations are treated in Section 4, where we show
how to generate periodic analogs of the known static axisymmetric
solutions; we give only a sketchy account of the non-static case,
however, as our results are still incomplete, requiring more
sophisticated techniques.
In Section 5, all these results are combined to derive
the new solutions with non-trivial topology. Section 6 is devoted
to a discussion of the new symmetries and the action of the
Virasoro-Witt algebra on moduli space.

\section{Flat metrics on Riemann surfaces}

Hereafter we denote by $\L$ a Riemann surface of genus $g$ with
local coordinates $(z,\zb)$. The metric on $\L$ may be chosen in many
different ways. This is related to a well known and basic property of
conformal field theories, including string theory, namely their
invariance with respect to local variations of the metric. An
especially convenient choice is based on the Mandelstam picture of
string interactions in the light-cone gauge \c{Man, GW}, where the
metric is taken to be flat and the unitarity of the theory is manifest.
Except for genus $g=1$, such a metric will unavoidably have
$\delta$-function singularities of the curvature, yielding the
standard Euler characteristic $\chi = 2-2g$ for a surface of genus $g$
(in the string theory interpretation, these singularities correspond
to points where strings split or join). Note that the singularities
can always be smoothed out by means of a conformal transformation of
the metric. By contrast, the models considered here, which are
obtained from a reduction of the Einstein equations in higher
dimensions, are {\it not} conformally invariant\footnote{At least not in
the usual sense. However, we would like to draw the reader's attention
to the fact that in many respects
the dimensionally reduced theory resembles Liouville
theory, where the conformal factor also does not decouple, but
conformal invariance can be restored nonetheless at the quantum level.
See also \c{Nic2} and Section 7
for a discussion.}. Thus, the singular points
cannot be eliminated here: this is the price we have to pay for
topological non-triviality of the solutions.

As already explained in the introduction, we pick a meromorphic
differential $d\xi(P)$ and define a flat metric on $\L$ by (\r{ds0}).
Away from the zeros and the poles of $d\xi(P)$ and in accordance
with (\ref{xiP}), we can adopt $\xi(P) = x(P) + i\rho (P)$
as a local coordinate on any simply connected
region of $\L$. Let us now discuss in turn the
various situations arising from the different choices of $d\xi(P)$.
We remark that among these, the case where $d\xi$ is a
differential of the second kind (i.e. having higher order poles)
is usually not considered in the string literature.

\subsection{$d\xi$ is a differential of the first kind}

The linear space of holomorphic differentials (abelian differentials
of the first kind) on $\L$ has complex dimension $g$. According to the
Riemann-Roch theorem, any holomorphic differential $d\xi$ on $\L$
possesses $2g-2$ zeros at some points $P_1,...,P_{2g-2}$; for
simplicity, we will assume that $d\xi$ has only simples zeroes.
Away from these points the curvature vanishes.
In the neighbourhood of any zero $P_j$, the
metric (\r{ds0}) is degenerate, and we have
\be
ds_0^2 =C |z|^2   dz d\zb\;,\;\;\;\;\; C>0
\;\;\;\; {\rm as} \;\;\;\; P \sim  P_j
\la{dsd}\ee
where $z(P)$ is a local coordinate in the neighbourhood of $P_j$
such that $z(P_j) =0$.
The Gaussian curvature of an arbitrary metric $g(z,\zb) dz d\zb$
has the form
$$ {\cal K} =- \f{(\log g)_{z\zb}}{2g} $$
and the Euler characteristic of $\L$ is defined by
\be
 \chi = \f{1}{2\pi} \int_{\L} g {\cal K} d z d\zb
\la{hi}\ee
Substituting the above form of the metric near $P_j$, we find
$$g {\cal K} = -\f{1}{2}
     (\log |z|^2)_{z\zb} =- 2\pi \delta(u)\delta(v), \;\;\;\;\;
\;\;u\equiv {\rm Re} z,\;\;\;\;\;v \equiv {\rm Im} z $$
and therefore the integral (\r{hi}) receives a contribution $(-1)$ from
each singular point; these contributions add up to the expected
result $\chi = 2 -2g$. Geometrically, the different flat pieces of
$\L$ are glued together at the ``interaction points'' $P_j$ \c{GW}.
Singularities are absent only for the
torus, for which $2-2g=0$; there is then only one holomorphic
differential $dz$. To be able to interpret these ``naked singularities''
and the singularities of the associated four-manifold in Section 3,
we introduce polar coordinates in the vicinity of $P_j$ by
$$ r=|z|\;\;\;\;\;\;\theta = \arg z $$
Defining a new radial coordinate $R=\f{\sqrt{C}}{2}r^2$,
the metric (\r{dsd}) near $P_j$ is cast into the form
\be
ds_0^2 =           dR^2 + 4 R^2 d\theta^2
\la{cs}\ee

\subsection{$d\xi$ is a differential of the second kind}

Next consider a meromorphic differential $d\xi$ on $\L$ with poles of
order $>1$ only. For simplicity, we assume that $d\xi$ has only
poles of order two located at the points $R_1,..., R_n$:
$$
d\xi (P) = C_j\f{dz}{z^2} +O(1),\;\;\;{\rm as}\;\;\;P\sim  R_j\;,\;\;\;
j=1,...,n    $$
Introducing a new complex coordinate $\zeta = - z^{-1}$ in the
neighbourhood of $R_j$, we get
$$ ds_0^2 =\tilde{C} d\zeta d\bar{\zeta}
         \;\;;\;\;\;\; \zeta\sim \i\;\;\;\;
{\rm as}\;\;\;\;\; P\sim  R_j $$
Consequently, a small disc around the point $R_j$ represents an
asymptotically flat region on $\L$; the number of such regions
is equal to $n$. The number
of zeroes of $d\xi$ on $\L$ is now equal to $2g-2+2n$; the
behaviour of the metric at these points was discussed in
the previous subsection. As a special case, we can have
$d\xi =d h(P)$, where $h(P)$ is some meromorphic function
on $\L$ (here with simple poles); then both functions
$x=\re \, h(P)$ and $\rho =\im \, h(P)$ are single-valued on $\L$.
We do not know whether differentials of the second kind
admit an interpretation in the context of string theory.

\subsection{$d\xi$ is a differential of the third kind}

Finally, consider abelian differentials of the third kind,
having first order poles at the points $Q_1,...,Q_n$
with purely imaginary residues (this
is sufficient for our purposes):
\be
d\xi (P) = \Big(- \f{i\alpha_j}{z} + O(1)\Big) d z\;\;\;,
       \;\;\;\alpha_j\in \R
\;\;\;\;{\rm as}\;\;\;P\sim Q_j\;;\;\;\;\sum_{j}\alpha_j =0
\la{res}\ee
For $\xi(P)$ close to $Q_j$ we have
$$ \xi(z)\sim - i\alpha_j\log z\;\;\;\; {\rm as}\;\;\;\;z\sim 0$$
whence
$$x\sim \alpha_j (\arg z + 2\pi k)\;\;\;\;\;\;\rho\sim\alpha_j
\log|z|$$
where $k\in {\bf Z}$.
Thus the small disc centered at $Q_j$ represents a semi-infinite
tube such that $\rho$ increases logarithmically
to infinity as one moves along the tube and $x$ parametrizes the
transverse direction (and is thus only defined modulo $2\pi \alpha_j$).

Relation (\r{res}) defines $d\xi$ only up to an arbitrary linear
combination of holomorphic differentials. However, if we impose the
additional restriction that all cyclic periods of $d\xi$ are {\it real}
(so there are $2g$ real conditions), it will be uniquely defined. In
this case $\rho \equiv \im \,\xi$ is a single-valued function on $\L$.
This is precisely the differential used in the Mandelstam
description of multiple string scattering in the light-cone gauge;
the globally defined coordinate $\rho$ plays the role of
light-cone time, and the multi-valued coordinate $x$ parametrizes
the various strings at any given time. Since the number of zeroes of an
arbitrary meromorphic differential is equal to the number of
its poles plus $2g-2$, $d\xi$ has altogether $2g-2+n$ zeroes on $\L$,
which correspond to the string interaction points as already explained.
The semi-tubes growing out of the surface near $Q_j$ are interpreted
as asymptotic in- or out-states of free strings, depending on whether
the residue is positive or negative (the residues are identified
with the momenta of the in- and outgoing strings, so the vanishing
of their sum expresses nothing but momentum conservation).

When discussing solutions of the Ernst equation on $\L$, we can, of
course, also allow for linear combinations of differentials of the
the second and third kinds.

\section{Simple examples of topologically non-trivial flat manifolds}

Before discussing the Ernst equation on $\L$, we find it instructive
to explain how our construction works for $\E=1$, leading to
topologically non-trivial flat manifolds $\M$ in four dimensions,
i.e. with metric
\be
ds^2 = d\xi d\xb + \rho^2 d\phi^2 - dt^2
\la{flat}
\ee
where, of course, the notation is the same as in the previous sections.
Since all examples are topologically the product of the time
axis $\R$ and a spatial flat three-manifold $\M_0$, we can effectively
ignore the time coordinate in the sequel.
Evidently, for the metric (\r{flat}) to be well defined, it
is necessary that $\rho (P)$ be globally defined. This
shows again that all cyclic periods of
$d\xi (P)$ should be real. Owing to the fact that the period
matrix associated with $\L$ has positive imaginary part, this
condition cannot be satisfied for holomorphic $d\xi(P)$.
Consequently, $d\xi(P)$ must have poles
of some kind. Its singular parts and the requirement of reality
determine $d\xi$ uniquely. We now discuss the properties
of $\M_0$ taking into account the possible choices of abelian
differentials $d\xi$.

Suppose that $d\xi(P)$ has a zero at $Q\in \L$. In the vicinity
of $Q$, the worldsheet metric can be brought into the form (\r{cs}),
so the associated metric on $\M_0$ becomes
\be
ds^2 = dR^2 + 4R^2 d\theta^2  + \rho^2 d\phi^2
\la{cs2}
\ee
where $\rho$ at $R=0$ is given by $\rho (Q)$, and $R$ and $\theta$
are local polar coordinates in the $\xi$-plane. If the second term
were $R^2 d\theta^2$, the space would obviously be flat.
On the other hand, if instead of $\beta^2 =4$, the prefactor obeyed
$\beta ^2 < 1$, the metric would correspond to a cosmic
string with deficit angle $2\pi (1-\beta)$ around $R=0$ \c{Vil}.
In the case at hand, there is an excess angle of $2\pi$ instead.
Accordingly, we identify this singularity with a cosmic string
of negative tension at $R=0$; since $0\leq \phi < 2\pi$, this
cosmic string forms a ring of radius $\rho (Q)$. By construction,
there are as many cosmic strings as there are zeroes of $d\xi$ in the
domain of positive $\rho$. Encircling such a string once, one
ends up in another (flat) axisymmetric ``universe'', while a rotation
of $4\pi$ will bring the observer back to the universe from where
he or she started; this is the physical meaning of the excess
angle. Analogously, the number of different ``universes'' that can
be reached in this fashion increases with the order of the zero
(it is easy to see that the excess angle is then $4\pi, 6\pi,...$),
and for each extra zero of $d\xi$, new separate ``universes'' become
accessible.

The properties of the metric (\r{flat}) near the poles
of $d\xi$ can be likewise
understood on the basis of the results of Section 2. For instance, let
$P_0$ be a second order pole of $d\xi$, and $z= z(P)$ a
local coordinate at $P_0$ defined in such a way that $d\xi\sim
z^{-2}dz$ as $P\sim P_0$. Clearly, $\xi=z^{-1} \rightarrow\i$ as
$P\rightarrow P_0$, implying
$\rho(z)=\im \, [z^{-1}]$ and $x(z)=\re \, [z^{-1}]$.
The small disc around the second order pole of $d\xi$ represents an
asymptoticaly flat region on $\M_0$ (the same conclusion holds
for poles of yet higher order). On the other hand, if
$P_0$ is a simple pole of $d\xi$ with imaginary residue,
a small disc on $\L$ centered at $P_0$ is
topologically equivalent to a semi-infinite tube.
The domain in $M_0$ corresponding to this tube is the
product of the domain $\rho > \rho_0$ in the $(\rho,\phi)$-plane
and a circle $S^1$ in terms of the  $x$-coordinate ($\rho_0$ is a
constant related to diameter of the disc). In other words, the
global structure of $\M_0$ can be described as follows.
If $\L$ were just the complex plane with coordinates $(x,\rho )$
we would get the space $\M_0 = \R^3$ by rotating the half-plane
$\rho \geq 0$ around the symmetry axis $\rho =0$. This procedure
may be repeated for arbitrary $\L$. Consider the part $\L^+$ of $\L$
for which $\rho \geq 0$; then $\M_0$ is obtained ``rotating'' $\L^+$
around the contour $\cal C$ consisting of all points $P\in \L$ such that
$\rho (P) =0$ (observe that $\cal C$ may have several disconnected
components). Rigorously speaking,
we get a foliation $\M_0\rightarrow \L^+$ where the fibre corresponding
to each point of $\L^+$ is a circle of radius $\rho$. The flat metric
is properly defined on all of $\M_0$ and regular away from the
poles and zeroes of $d\xi$ as we already explained.
The total number of cosmic strings, asymptotic
regions and semi-infinite tubes is constrained by the Riemann-Roch
theorem according to section 2.

Notice that in our construction the subset $\L_- \subset \L$
corresponding to $\rho < 0$ can be ignored, because we admit only
non-negative $\rho$ in order to be able interpret the metric
as a genuine metric in four dimensions.
Taking into account the reflection symmetry
$\rho\rightarrow -\rho$ on $\M$
of the metric (\r{flat}), the section $\phi = k\pi,\;\; k=1,2$
of $\M_0$ may be identified with $\L^+$ glued to its mirror image
along the ``symmetry axis'' $\rho =0$. So without loss of generality
we could have assumed from the beginning that the curve $\L$ admits an
anti-holomorphic involution $\tau :\L\rightarrow \L$, $\tau^2 =1$
such that $\tau(\L^+) =\L^-$ and could have defined $\cal C$
as the set of points invariant with respect to $\tau$
(locally, the involution $\tau$ maps $(x,\rho)$ to $(x,-\rho)$).
In this case the section
$\phi=k\pi\;,\;\;k=0,1$ of $\M_0$ may be identified with the surface
$\L$ itself.

The above considerations can be extended in an obvious manner to
the simpler case of three-manifolds with metrics
\be
ds^2 = d\xi d\xb + dx_3^2
\la{fm} \ee
whose third coordinate $x_3$ is unrelated to $\xi(P)$; from the
two-dimensional point of view, we are dealing here with a
constant dilaton. The essential difference is that we now
need not require real periods. Topologically, $\M_0$ is then
a product $\L \times {\bf R}$ or $\L \times S^1$.

In a somewhat different context, flat Lorentzian
three-manifolds having {\it non-singular}
space-like surfaces at any finite value of
the time parameter, but initial and final singularities
at $t=0$ and $t=\infty$, were constructed in \c{Witten}.

\section{Periodic solutions of the Ernst equation}

Since our construction requires periodic solutions of the
Ernst equation, we will in this section explicitly demonstrate
that a large class of solutions actually exists (the periodic analog
of the Schwarzschild solution was already constructed in \c{Myers}).
While so far we have explicit examples only for the static case
(i.e. $A=0$ in (\r{dL})), we give arguments why periodic solutions
{\it with rotation}, and in particular a periodic analog of the Kerr
solution, should also exist; however, their explicit construction
requires more sophisticated tools
from the theory of integrable systems.

\subsection{Static case}

As is well known, static metrics with $A=0$ correspond to real Ernst
potentials. For these, the Ernst equation (\ref{E1}) can be
linearized by the substitution
\[ \o =\log \E \]
and is thereby reduced to the Euler-Darboux equation
\be
\o_{xx} +\f{1}{\rho}\o_{\rho} +\o_{\rho\rho} =0
\la{ED}\ee
The metric (\ref{ds}) becomes
\be
ds^2 =e^{-\o}\big[e^{2k}(dx^2 +d\rho^2)
                                      +\rho^2
                                            d\phi^2\big] + e^{\o} dt^2
\la{m1}\ee
where the conformal factor is determined from the first order equation
\[ k_{\xi} =\f{i\rho}{2}(\o_{\xi})^2 \]
or, equivalently,
\be
k_{\rho}=\f{\rho}{4} \big( \o_{\rho}^2 -\o_x^2\big)\;\;\;\;\;\;\;
k_x=\f{\rho}{2}\o_x\o_{\rho}
\la{cfac}\ee
We can now construct $x$-periodic analogs of known static
solutions by means of the following procedure.

Let $\o_0(x,\rho)$ be some solution of (\r{ED}).
Consider the expression\footnote{The following construction is
inspired by the construction of the Weierstrass
$\wp$-function.}
\be
\o(x,\rho)=\sum_{n=-\i}^{\i}\Big\{\o_0(x+nL,\rho)+a_n\Big\}
\la{ps}\ee
where the coefficients $a_n$ are constants to be chosen in such a way
that series (\r{ps})  becomes convergent. It is important that these
coefficients do not depend on $(x,\rho)$ since otherwise the sum
could not possibly satisfy the original equation (\r{ED}). If
convergence may be achieved, the function (\r{ps})  describes a
solution of (\r{ED}) with period $L$. Clearly, in order
to obtain a truly periodic metric (\ref{m1}) we must also verify the
periodicity of the function $k(x,\rho)$ defined by (\ref{cfac}).

The following theorem shows that our method of construction
is quite general.

\begin{theorem}
Let $\o_0(x,\rho)$ be any solution of the Euler-Darboux equation
corresponding to an asymptotically flat metric (\ref{m1}), i.e.
\be
\o_0(x,\rho)=\f{\beta}{r} +O(r^{-2})\;\;\;\;\;{\rm as}\;\;\;\;\;
r\rightarrow\i
\la{af}\ee
where $r=\sqrt{x^2 +\rho^2}$; $M=-\f{1}{2} \beta$ is the mass. Let
\be
a_n = -\f{\beta}{L|n|}\;\;,\;\;\;n\neq0\;\;,\;\;a_0=0
\la{an}\ee
Then series (\ref{ps}) is convergent for all $(x,\rho)$ except
the points $(x_0+nL,\rho_0)$, where the function
$\o_0(x,\rho)$ is singular ($n\in {\bf Z}$), and defines a periodic
function with period $L$.
\end{theorem}

{\it Proof:} For large $n$ we have
\[ \o_0(x+nL,\rho) + a_n = \beta\Big(\f{1}{\sqrt{(x+nL)^2+\rho^2}} -
\f{1}{L|n|}\Big) + O\big(\f{1}{n^2}\big) = O\big(\f{1}{n^2}\big) \]
by (\ref{af}), and therefore the series (\ref{ps}) converges if
$(x+nL,\rho)$ does not coincide with a singular point of $\o_0(x,\rho)$
for any $n$. $\Box$

So starting from an arbitrary static asymptotically flat solution
we can construct its $x$-periodic analog. Consider for instance
the Schwarzschild solution,
which is characterized by the Ernst potential
\be
\o_0=\log\E_0 \;\;\;\;\;\;\;
\E_0 (x,\rho ) =\f{\s{(x-M)^2+\rho^2} + \s{(x+M)^2+\rho^2} -2M}
{\s{(x-M)^2+\rho^2} + \s{(x+M)^2+\rho^2} +2M}
\la{S}\ee
where $M \in {\bf R}$ is an arbitrary positive constant (the mass
of the black hole). Here, all square roots are taken to be positive;
this means that we do not consider (\ref{S}) inside the event horizon,
which coincides with the segment $\rho=0\,,\,x\in [-M,M]$.
The coefficient $\beta$ in (\ref{af}) is therefore
equal to $-2M$. Thus the periodic analog of the
Schwarzschild solution (\ref{S}) has the following form:
\be
\E(x,\rho)=\E_0(x,\rho)\prod_{n=1}^{\i}\E_0(x+nL,\rho)\E_0(x-nL,\rho)
\exp\Big( \f{4M}{nL} \Big)
\la{pS}\ee
Obviously,
\[ \E(x+L,\rho)=\E(x,\rho) \]
This is essentially the solution found in \c{Myers}.
In the sequel we will assume $\f{1}{2} L > M$ (for $L\leq 2M$,
the interpretation of the solution is not clear
as the horizon overlaps with itself, and the Ernst potential
vanishes on the symmetry axis). Convergence
of the infinite product (\ref{pS}) is equivalent to convergence of the
series (\ref{ps}) and thus guaranteed by Theorem 1 (for non-negative
values of the square roots in (\ref{S})). Consequently, the
solution (\ref{pS}) is a periodic function on the upper half plane
$\rho \geq 0$ with ``fundamental region''  $\cal F$ defined by
$\rho \geq 0\,,\,-\f{1}{2} L\leq x \leq\f{1}{2} L$.

It is now not difficult to verify that the
function $\E(x,\rho)$ defined by (\ref{pS}) is smooth
everywhere on $\cal F$ away from the points $x=\pm M\,,\,\rho =0$.
and non-zero everywhere except on the horizon
(i.e. $\rho =0 \,,\, \x \leq M$). The periodicity of the
conformal factor, i.e.
\be k(x+L,\rho)= k(x,\rho) \la{kp} \ee
is implied by
\[ \int_{-L/2}^{L/2} k_x dx = 0 \]
where the derivative $k_x$ is to be evaluated by means
of (\ref{cs}). This, in turn is a consequence of the vanishing
of the following contour integral
\be
\int_{l} \Big\{\f{\rho}{4}(\o_{\rho}^2 -\o_x^2)d\rho +\f{\rho}{2}
\o_x\o_{\rho} dx\Big\}
\la{ci}\ee
where the closed contour $l$ is depicted  in Fig.1. This
integral vanishes because the function $\o$ obeys (\ref{ED}) and is
smooth everywhere inside of $l$. The integrals along the edges
$[(-\Lh,0),\;(-\Lh,\rho)]$ and $[(\Lh,\rho),\;(\Lh,0)]$ cancel
due to the periodicity of $\o(x,\rho)$. Owing to the presence of the
factor $\rho$ in (\ref{ci}) the contribution of the interval
$[(\Lh,0),\;(-\Lh,0)]$ reduces to a sum of contributions of two small
rectangular paths around the points $x=-M$ and $M$ (cf. Fig.1.),
where the derivatives $\o_x$ and $\o_{\rho}$ become singular. These
contributions cancel by virtue of the symmetry
\[ \o(-x,\rho) =\o(x,\rho) \]
inherited by solution $\E$ from $\E_0$. So the integral along the
contour $[(-\Lh,\rho),\;(\Lh,\rho)]$ also vanishes and we get
$k(-\Lh,\rho)=k(\Lh,\rho)$. In conclusion, the metric
(\ref{m1}) corresponding to the periodic Ernst potential
(\ref{pS}) is also periodic.

The asymptotic behavior of the Ernst potential (\ref{pS}) is given by
\be
\E=C\rho^{4M/L} \Big( 1+o(1)\Big)\;\;\;{\rm as}\;\;\;\rho\rightarrow\i
\la{ass}\ee
where $C$ is some constant. To see this, recall that the
function $\o=\log\E$ is defined by
\[ \o(x,\rho)=\o_0(x,\rho) + \sum_{n=1}^{\i}\Big[\o_0(x+nL,\rho) +
\o_0(x-nL,\rho) +\f{4M}{nL}\Big] \]
Substituting the explicit expression for $\o_0(x,\rho)$ (\ref{S})
and differentiating with respect to $\rho^2$, we obtain
\[ \frac{\partial\o}{\partial(\rho^2)} (x,\rho)=
\sum_{n=-\i}^{\i} \f{2M \big[s_1(n) + s_2(n) \big]}{\big[
   s_1(n)+s_2(n)+2M\big]\big[s_1(n)+s_2(n)-2M\big]
s_1(n) s_2(n)} \]
where $s_1(n)=\s{(x+nL+M)^2+\rho^2}$ and
$s_2(n)=\s{(x+nL-M)^2 +\rho^2}$.
The leading term in this series for large $\rho$ can be estimated
by approximating the sum by an integral; it is given by
\[ \sum_{n=-\i}^{\i}\f{M}{((x+nL)^2 +\rho^2)^{3/2}} =
       \f{2M}{L}\f{1}{\rho^2} \Big( 1+o(1) \Big) \]
Thus,
\[ \o = \f{2M}{L}\log\rho^2 + O(1)\;\;\;\;\;\;{\rm as}
\;\;\;\;\rho\rightarrow \i \]
and $\E = C\rho^{4M/L} (1+o(1))$ for some constant $C$.
Hence, as $\rho \rightarrow \infty$, the metric
(\ref{m1}) tends to the Kasner solution
\be
ds^2= \tilde C \rho^{\f{\alpha^2}{2} -\alpha} (dx^2 +d\rho^2)
   + C^{-1} \rho^{2-\alpha}
d\phi^2 - C \rho^{\alpha} dt^2
\la{as}\ee
where $\tilde C$ is another constant of integration and
the Kasner parameter $\alpha$ is related to the period $L$ by
$\alpha=4M L^{-1}$, so that $0\leq \alpha <2$ with our assumption
on the range of $M$.

The solution (\r{pS}) has a compact event horizon coinciding with the
segment $\rho=0\,,\,-M\leq x\leq M$. Outside the horizon it is
everywhere non-singular, including the segment of the symmetry axis
outside the horizon. Using the standard product representation
for the $\Gamma$-function, we find
\be
\E(x, \rho =0)= \exp \Big( \f{4\gamma M}{L} \Big) \,
\f{ \Gamma \Big( \f{\x +M}{L} \Big)
    \Gamma \Big( 1-\f{\x -M}{L} \Big)}
  { \Gamma \Big( \f{\x -M}{L} \Big)
    \Gamma \Big( 1-\f{\x +M}{L} \Big)}
\la{sa}\ee
for $M \leq \x \leq \f{1}{2} L $ ($\gamma$ is the Euler
constant), and $\E \equiv 0$ for $\rho =0$ and $\vert x \vert \leq M$.
As a consequence of the reflection and translation symmetry, the
free integration constant in equation (\ref{cs})
may be chosen in such a manner that conical singularities on the
part of the symmetry axis outside the horizon are avoided
(this requirement fixes the constant $\tilde C$ in (\ref{as})).
In the limit $L\rightarrow\i$, the solution obviously tends (pointwise)
to the ordinary Schwarzschild solution. The leading term in the
asymptotic expansion then approaches the flat metric, as it
should be\footnote{Alternatively, one could
regard this solution as describing an
infinite chain of black holes spaced at a distance $L$. At first sight,
it seems remarkable that this configuration does not require conical
singularities on the axis between adjacent black holes for
stability. Rather, it appears to be stabilized by its symmetry under
reflections and translations and the presence of infinitely many
black holes ``on each side''. However, this also indicates instability
under non-periodic perturbations, which makes this interpretation
somewhat less attractive.}.

By theorem 1 we can obtain a periodic counterpart of any
asymptotically flat static solution of (\r{E1}). The method described
in this section fails, however, for non-static axisymmetric
stationary solutions, because the Ernst equation can no longer be
linearized in this case. Since the extension requires entirely
new techniques, we shall sketch a possible method based on methods
borrowed from soliton theory in the next section.

\subsection{General case}

The Ernst equation (\r{E1}) may be derived as the compatibility
condition of the following linear system \c{BZ, M, N} (for more
recent developments, see \c{BM, Nic1}, and \c{KM}, whose notation
and conventions we will follow in this section):

$$ \Psi_{\xi}= U\Psi\;\;\;\;\;\;\;\;\;\Psi_{\xb}=V\Psi $$
\be\la{ls}\ee
$$   U=    \left(\ba{cc} C_1\;\;\;\;\;0\\
                               0\;\;\;\;\;C_2\ea\right)+
\sqrt{\f{\l-\xb}{\l-\xi}}
\left(\ba{cc}0\;\;\;\;\;C_1\\
             C_2\;\;\;\;\;0\ea\right)\;\;\;\;\;\;\;\;\;
V  =\left(\ba{cc} C_3\;\;\;\;\;0\\
                               0\;\;\;\;\;C_4\ea\right)+
\sqrt{\f{\l-\xi}{\l-\xb}}
\left(\ba{cc}0\;\;\;\;\;C_3\\
             C_4\;\;\;\;\;0\ea\right)  $$
where $\l\in \C$ is the spectral parameter\footnote{The spectral
parameter $\lambda$ is called $w$ in \c{BM,Nic1}. Sometimes it is
convenient to use the $(x,\rho)$-dependent spectral
parameter $t = \f{1}{\rho}(\l -x+\sqrt{(\l -x)^2 +\rho^2 }$
(first introduced in \c{BZ}) mapping $\S$ to the complex plane.};
$\Psi (\l,\xi,\xb)$ is
a two-by-two matrix, whose coefficients $C_j (\xi,\xb)$
are related to the Ernst potential $\E$ by
\be
C_1=\f{\E_{\xi}}{\E+\Eb}\;\;\;\;\;\; C_2=\f{\Eb_{\xi}}{\E+\Eb}\;\;\;\;\;
C_3=\f{\E_{\xb}}{\E+\Eb}\;\;\;\;\;\; C_4=\f{\Eb_{\xb}}{\E+\Eb}\;\;\;\;\;
\la{ABC}\ee
The identification (\r{ABC}) is a consequence of the asymptotic
condition
\be
  \Psi(\l=\i)=\left(\ba{cc} 1\;\;\;\;\; \E \\
                           -1\;\;\;\;\; \Eb \ea\right)
\la{nc}\ee

Considered as a function of the complex spectral parameter $\l$
with fixed $(\xi,\xb)$, the linear system
(\r{ls}) lives on the two-sheeted Riemann surface $\S$ of
genus zero defined by the equation

$$ \gamma^2 =(\l-\xi)(\l-\xb);$$
$\Psi$ is thus a function of the three variables $(P,x,\rho)$, where
$P\in \S$ (actually, we should distinguish points on the spectral
curve $\S$ from points on the worldsheet $\L$,
but we will nevertheless use letters
$P,Q,...$ if there is no danger of confusion). By $\sigma$,
we denote the involution on $\S$ interchanging the two sheets.

The existence of the linear system (\r{ls}) permits us to construct
solutions of (\r{E}) by means of the inverse
scattering method \c{BZ}-\c{Kor},\c{G2},\c{BM}.
So far, most efforts in this direction
have been concerned with the investigation of asymptotically
flat solutions of (\r{E1}). Although our results
are still incomplete, we now present some evidence that
every asymptotically flat solution of (\r{E1}) should
have a natural periodic counterpart.

To illustrate these arguments,
let us attempt to construct the periodic analog of the
Kerr solution. For this purpose, recall the standard
construction of $2n$-soliton solutions of (\r{E1}) (there are no
solutions with an odd number of solitons). The associated
function $\Psi_n(P)$ has the following form:
\be
\Psi_n(P)= \f{C(\xi,\xb)}{\l^n}
           \left(\ba{ccc}  Q_n(\l)\;\;\;\;\;\;\gamma P_{n-1}(\l) \\ \\
               \gamma \ol{P_{n-1}(\bar{\l})}\;\;\;\;\;\;\;
                                            \ol{Q_n(\bar{\l})} \ea
                                               \right)
\left(\ba{ccc}1\;\;\;\;\;\;1\\   \\
         -1\;\;\;\;\;\; 1 \ea\right)
\la{nkerr}\ee
where
$$Q_n(\l)=\sum_{k=1}^{n} q_k\l^k\;\;\;\;\;\;\;\;
P_{n-1}(\l)=\l^{k-1}+\sum_{k=1}^{n-2} p_k\l^k $$
are two polynomials with $(x,\rho)$-dependent coefficients
$p_k$, $q_k$; $C(\xi,\xb)$ is a diagonal matrix providing normalization
of the first column according to (\r{nc}): $C_{11}=(q_n-1)^{-1}$;
$C_{22}=(1-\bar{q}_n)^{-1}$.
             For arbitrary coefficient functions $p_k$ and
$q_k$, the function $\Psi_n$ will
not satisfy the linear system (\r{ls}) because its
logarithmic derivatives
$\Psi_{\xi}\Psi^{-1}$ and $\Psi_{\xb}\Psi^{-1}$ will have not only
the poles at $\l=\xi$ and $\l=\xb$ required by (\r{ls}),
but also at the zeroes
$\l_1,...\l_{2n}$ of $\det\Psi_n$. To eliminate these additional poles
suppose that the points $\l_1,...,\l_{2n}$ are independent of
$(\xi ,\xb)$, and that the eigenvectors of $\Psi_n$ at
these points are also $(\xi,\xb)$-independent, i.e.
\be
\Psi_n(\l_k)\left(\ba{cc} 1\\ d_k \ea\right) =0 \;\;\;\;\;k=1,...,2n
\la{reg}\ee
where $d_k$ are some constants. From (\r{reg}) we can deduce
$4n$ linear equations for the $2n$ functions
$p_0,...,p_{n-2}$, $q_0,...,q_{n}$. To ensure their solvability
it is enough to require that
\be
\l_{2k-1}=\bar{\l}_{2k-1} ,\;\;\;\;\;\;\l_{2k}=\bar{\l}_{2k} \;\;\;\;\;
d_{2k-1}=-\bar{d}_{2k-1} \;\;\;\;\;\;d_{2k}=-\bar{d}_{2k}
\la{rea1}\ee
or
\be
\l_{2k-1}=\bar{\l}_{2k}  \;\;\;\;\;\;\; d_{2k-1}=-\bar{d}_{2k}
\la{rea2}\ee
for $k=1,...,n$.

The related Ernst potential is then easily identified as
$$\E= \f{q_n+1}{q_n-1}$$
where $q_n$, by Kramer's rule, is the ratio of two $2n\times 2n$
determinants. This expression is more conveniently dealt with
in terms of the potential
$$\Gamma \equiv\f{1-\E}{1+\E}= -\f{1}{q_n} $$
We will not write down the corresponding expressions for
arbitrary $n$; they can be found in \c{ExactSol}.

For $n=1$ and $\l_{1,2}\in {\bf R}$, $d_{1,2}\in i{\bf R}$ we get
\be
\Gamma^{-1} = \f{c_1-c_2}{2} X + \f{c_1 +c_2}{2} Y
\la{G}\ee
where
$$c_j \equiv\f{-1 +d_j}{1+ d_j}\;,\;\;\;\;j=1,2 $$
are new arbitrary constants satisfying $|c_j| =1$, and
$$ X\equiv\f{1}{\l_1-\l_2}\Big[\sqrt{(\l_1-\xi)(\l_1-\xb)}+
\sqrt{(\l_2-\xi)(\l_1-\xb)} \Big] $$
$$ Y  \equiv\f{1}{\l_1-\l_2}\Big[ \sqrt{(\l_1-\xi)(\l_1-\xb)}-
\sqrt{(\l_2-\xi)(\l_1-\xb)} \Big]  $$
are prolate ellipsoidal coordinates. This is nothing but the
well known Kerr-NUT solution; to get the Kerr solution itself, we
only have to put $c_2=-\bar{c_1}\equiv c$. To recover the
multi-Kerr solution with arbitrary parameters, we have
to choose $c_{2j-1}=-\bar{c}_{2j}$,  $|c_{2j-1}|=|c_{2j}|=1$
for all $j=1,...,g$.

Now let us turn to possible periodic generalizations.
Intuitively, one would expect a periodic analog of the Kerr solution
to represent an infinite superposition of identical Kerr black holes
lined up on the symmetry axis at equal distances. Analytically,
for an arbitrary solution of (\r{E1}) obeying the
periodicity condition (\r{per}), the associated linear system matrices
$U$ and $V$ from (\r{ls}) must satisfy the relations
$$ U(\l+L,x+L,\rho)=U(\l,x,\rho)\;\;\;\;\;\;\;\;
   V(\l+L,x+L,\rho)=V(\l,x,\rho)    $$
Then for a solution $\Psi$ of (\r{ls}) we have
\be
\Psi(\l+L,x+L,\rho) =\Psi(\l,x,\rho) R(\l)
\la{PP}\ee
where $R(\l)$ is some $(x,\rho)$-independent matrix.
We emphasize that the world-sheet coordinate $x$ and the spectral
parameter $\l$ must be shifted simultaneously by the same amount in
order to maintain the $\l$-dependence prescribed by the linear system
(\r{ls}). Note also that, for topologically non-trivial
worldsheets $\L$, $\Psi$ is {\it not} single-valued on $\L$ in general
even for $R(\l)=1$.

{}From (\r{PP}) we infer that the $(x,\rho)$-independent parameters
in $\Psi$ must be invariant under the transformation
$\l\rightarrow\l+L$ (in the general multi-soliton case, these
parameters include the positions of the points
$\l_j$ and the constants $c_j$). To construct a periodic analog
of the Kerr solution we thus have to start from
an infinite-soliton solution where the zeroes of $\det\Psi$ are
located at the points $\l_1+nL$ (with the same parameter $c_1$) and
$\l_2+nL$ (with the same paramter $c_2=-\bar{c_1}$). Just as for
the periodic Schwarzschild solution, we can avoid ``horizon
overlap'' by imposing the condition
$L>2M$, where $2M\equiv |\l_1 -\l_2|$.

A rigorous way to construct such a solution
would be to consider a sequence of Ernst potentials
$\E_n(x,\rho)$ , $n=1,2,...$, where $\E_n$ is the $4n+2$-soliton
solution of (\r{E1}) corresponding to the zeroes of $\det\Psi$ at
$\l_1+nL$ with the ``dressing parameters'' $c$ ($|c|=1$) and
$\l_2+nL$ with the ``dressing parameters'' $-\bar{c}$.
If we could prove the pointwise convergence of the sequence $\E_n$,
the limit solution $\E = \lim_{n\rightarrow \infty} \E_n$
could be regarded as a periodic
analog of Kerr solution. We have at this time no rigorous proof
of convergence, but can offer the following arguments in favor
of our hypothesis. First of all, the sequence is certainly
convergent when $c_2=-c_1=1$, in which case the ``dressing procedure''
would reproduce the periodic analog of the Schwarzschild solution
derived in the foregoing section. Secondly, for large
$\rho$, the solution should describe the gravitational field of an
infinitely extended rotating cylinder on the symmetry axis,
a solution which is known to exist \c{ExactSol}
(the periodic Schwarzschild
solution likewise approaches the field of an infinite static
cylinder). Moreover, the contribution from angular momentum
to the asymptotic form of the Ernst potential at $\rho \rightarrow
\infty$ is down by one order in comparison with the contribution
of mass (i.e. of order $O(\rho^{-2})$ rather than $O(\rho^{-1})$),
so they would not affect the convergence of the series.
Indeed, these arguments suggest that the solution with rotation
again approaches the Kasner solution (\r{as}) with the same
Kasner parameter. However, even assuming convergence of the
sequence $\E_n$, it is not clear at the moment whether the
``dressing procedure'' will not generate
unphysical singularities outside the event horizon
$[\l_1+nL,\;\;\l_2+nL]$ unlike in the static case, where such
additional singularities could be shown to be absent.

Analogously, in order to construct
periodic analogs of the algebro-geometrical
solutions of (\r{E1}), which contain the multi-soliton solutions as a
degenerate partial case \c{KM,Kor}), one would have to start from a
hyperelliptic algebraic curve of infinite genus with a periodic
configuration of immovable branch cuts.

The general conjecture is that every asymptotically flat solution
of the Ernst equation with mass $M$ has a natural periodic
analog which is asymptotically Kasner with Kasner parameter $4M/L$.
The spectral data of the fully periodic solution are obtained
by shifting the spectral data of the
initial solution by $nL$ for $n\in {\bf Z}$.
In the limit $L\rightarrow\i$ the periodic solution tends
(pointwise) to its asymptotically flat counterpart.
Assuming our conjecture to be true, the same correspondence
should hold between the symmetry groups pertaining to asymptotically
flat and periodic solutions, respectively,
i.e. there should exist a ``periodic Geroch group'', also
possessing a central extension. The corresponding
Lie algebras should, of course, be the same (i.e. coincide with the
affine Kac Moody algebra $A_1^{(1)}$).

Note that the periodicity constraint (\r{PP}) for $\Psi$
implies the periodicity of all related quantities depending either on
only $\lambda$ or on only $(\xi,\xb)$ (we can arrange $R=I$ by
a suitable transformation $\Psi\rightarrow
\Psi C(\lambda)$). For instance, with $F(\l)=I$ in (\r{nc}),
we get $\Psi(\l=\i,\xi+L)=\Psi(\l=\i ,\xi)$, which implies
periodicity of Ernst potential, i.e. $\E(x+L,\rho)= \E(x,\rho)$.
Besides, it is easy to check that
$$ M(\l)=\f{1}{\E+\Eb}\Psi^{t}(\l^{\sigma})\sigma_1 \Psi(\l) $$
(called ``monodromy matrix'' in \c{BM}) is always
$(\xi,\xb)$-independent. Hence, $M(\l)$ is periodic with period $L$
for $\Psi$ subject to (\r{PP}) with $R(\l)=I$.

As an example, we calculate the monodromy matrix for the periodic
Schwarzschild solution. The monodromy matrix associated with the
ordinary Schwarzschild solution (\r{S}) has the form
\be
T_0=-\f{(\l-M)(\l+M)}{\l^2}\sigma_3
\la{T0}\ee
where we put $\l_1=-M$, $\l_2=M$ and
$\sigma_3$ is the usual Pauli matrix. Now, for any static solution
(i.e. $\E=\Eb$), the related solution of the
linear system (\r{ls}) may be represented as
$$ \Psi(P)=\left(\ba{cc} \psi(P)\;\;\;\;\;\;\psi(P^{\sigma})\\
-\psi(P)\;\;\;\;\;\;\;\psi(P^{\sigma})\ea\right)
$$
where $\psi(P)$ is a solution of
$$(\log\psi)_{\xi}=\f{1}{2}(\log\E)_{\xi}
     \Big(1+\sqrt{\f{\l-\xb}{\l-\xi}}\Big)
\;\;\;\;\;\;\;
(\log\psi)_{\xb}=\f{1}{2}(\log\E)_{\xb}
      \Big(1+\sqrt{\f{\l-\xi}{\l-\xb}}\Big)$$
The related monodromy matrix is
$$ T=-\f{\psi(P)\psi(P^{\sigma})}{2\E}\sigma_3 $$

If $\E_1$ and $\E_2$ are two different solutions of (\r{E1})
with associated functions $\psi_1(P)$ and $\psi_2(P)$, and
monodromy matrices $T_1(\l)$ and $T_2(\l)$, respectively,
then $\E_1\E_2$ is a solution of (\r{E1}) corresponding to
$\psi(P)=\psi_1(P)\psi_2(P)$ with monodromy matrix
$T(\l)=T_1(\l)T_2(\l)$. From this simple observation, we can
derive the monodromy matrix corresponding
to the solution (\r{pS}) as an infinite product
$\prod_{n=-\i}^{\i}T_0(\l+ nL)$. Using the explicit form
(\r{T0}) of the matrix $T_0$, we obtain
$$ T(\l)=-\f{\sin\f{\pi(\l-M)}{L}\sin\f{\pi(\l+M)}{L}}
{\sin^2\f{\pi \l}{L}} \sigma_3 $$
In the different parametrization adopted in \c{BM}, we would get
$T(\l)={\rm diag}(r(\l),\;r^{-1}(\l))$ instead, where
$r(\l)=\sin\f{\pi(\l-M)}{L}\sin^{-1}\f{\pi(\l+M)}{L}$.
We observe the following difference between periodic and asymptotically
flat solutions: in the asymptotically flat case,
the dependence of the monodromy matrix on $\l$ can be inferred from
the knowledge of the Ernst potential on the symmetry axis \c{BM}.
Comparison of the monodromy matrix for the periodic Schwarzschild
solution with the related Ernst potential on the symmetry axis
(cf. (\r{sa}))  shows that this is no longer true
for periodic solutions.

\section{The Ernst equation on $\L$}

We will now combine the results arrived at
in the foregoing sections. That is, we will replace the
flat metric (\r{flat}) by the full metric (\r{ds}), and at the
same time assume the underlying world-sheet to be a non-trivial
Riemann surface, thereby introducing a non-trivial gravitational
field on the manifold $\M$. The non-flat metric on the
background provided by (\r{flat}) thus has the usual form of
the metric on a stationary axisymmetric Einstein manifold, viz.
\be
ds^2 =f^{-1} e^{2k} d\xi d\xb +f^{-1} (\im\xi)^2 d\phi^2
-f(dt +Ad\phi)^2
\la{met}\ee
where the functions $f,\;k$ and $A$ depending only on $(\xi,\xb)$
may be expressed in terms of the complex-valued Ernst potential
according to (\r{co}). The Einstein equations for the metric (\r{met})
reduce to the Ernst equation (\r{E1}), together with (\r{boxrho})
and (\r{co}), as already mentioned in the introduction.

Let us denote the poles of $d\xi$ on $\L$ with non-zero residues
by $Q_1,...,Q_n$. The reality of all cyclic
periods of $d\xi(P)$ implies that the related residues are
imaginary; as in section 2.3 we denote them by
$-i\alpha_j$ ($\alpha_j\in \R$). With
$(a_j,\;b_j)$ , $j=1,...,g$ the usual basis of homology cycles
on $\L$, the cyclic periods of $d\xi$ are given by
\be
A_j = \oint_{a_j} d\xi\;\;\;\;\;\;\;\;
B_j = \oint_{b_j} d\xi\;\;\;,\;\;j=1,...,g\;\;\;\;\;
2\pi\alpha_j=\oint_{C_j}d\xi\;\;\;,\;\;\;j=1,...,n
\la{peri}\ee
where $C_j$ are small contours enclosing the points $Q_j$. According to
our assumption, all these periods are real; therefore the function
$\rho(P)=\im\, \xi(P)$ is single-valued on $\L$, and $x(P)=\re\, \xi(P)$
has cyclic periods coinciding with (\r{peri}).

Suppose now that
$\E(x,\rho )$ is some solution of the Ernst equation. For
$\E(x,\rho)$ to be single-valued on $\L$, the following
periodicity conditions must be satisfied
$$\E(x+A_j ,\rho)=\E(x+ B_j,\rho)=\E(x,\rho)\;\;\;\;\;\;j=1,...,g$$
$$\E(x+2\pi \alpha_j,\rho) =\E(x,\rho)\;\;\;\;\;j=1,...,n$$
Since it is impossible to construct a continuous non-constant
function on $\R$ having two real periods $L_1$ and $L_2$ such that
the ratio $L_1/L_2$ is irrational, we must require that all cyclic
periods $A_j$, $B_j$ and $2\pi \alpha_j$ satisfy
\be
2\pi\alpha_j = k_j L\;\;\;\;\; j=1,...,n
\la{q1}\ee
\be
A_j= m_j L\;\;\;\;\;\;\;B_j=n_j L\;\;\;\;\;j=1,...,g
\la{q3}\ee
for some $L\in \R$ and certain integers $k_j, m_j, n_j$. If
$\E(x,\rho)$ is periodic with period $L$,
\be
\E(x+L,\rho) =\E(x,\rho)
\la{per}\ee
the function $\E \big( x(P),\rho (P) \big)$ is single-valued on
all of $\L$. The periodicity of the full metric (\r{met}) is
ensured if the same periodicity constraints are obeyed
by the functions $k$ and $A$, i.e.
\be
k(x+L,\rho) = k(x,\rho)\;\;\;\;\;\;
A(x+L,\rho)=A(x,\rho)
\la{perk}\ee
At least for the static case, these relations were established
in the previous section. Now we recognize that it is precisely
the periodicity constraints (\r{q1}) and (\r{q3}) which lead to
the discretization of moduli space.

Physically we can distinguish two essentially different situations:
\begin{enumerate}
\item   For $k_j=m_j=n_j=0$,
all cyclic periods of $d\xi$ vanish, and
$d\xi(P)=dh(P)$ for some meromorphic function $h(P)$ on $\L$; there
are then no restrictions on the surface $\L$. The condition (\r{per})
becomes empty, and $\E(h,\bar{h})$ solves (\r{E1}) without
further ado. In particular, we can choose $\E(x,\rho)$ to be
asymptotically flat (corresponding to the Schwarzschild solution, say);
then the metric (\ref{met}) will also be asymptotically flat near
the poles of $d\xi(P)=dh(P)$ (which are of second order at least),
where $x,\rho \rightarrow\i$.
The number of black holes on the symmetry axis
$\rho=0$ coincides with the number of poles of $h(P)$.
\item  If not all $k_j,m_j,n_j$ in (\r{q1}),(\r{q3}) vanish, the
differential $d\xi$ may in principle have poles
of arbitrary order. For definiteness, let us restrict attention
to the $L$-periodic Schwarzschild solution derived above.
The symmetry axis $\rho =0$ now consists of several disconnected
components. These may be either homeomorphic to $S^1$ or
non-compact. The first possibility corresponds to the cross
sections of $\L$ at ``time'' $\rho =0$. The second, on the other hand,
is realized in the vicinity of any higher order pole of $d\xi$, when
the pole is approached in such a way that $\rho=0$ but
$x\rightarrow\infty$. In the latter case, we end up with an
infinite number of black holes in this region, which would have been
asymptotically flat for $\E=1$. If one wants to avoid the occurrence
of infinitely many black holes, one must consequently assume that
$d\xi$ has only simple poles (which is precisely the situation
studied in string theory \c{GW}). The neighbourhoods of the simple
poles of $d\xi$ represent semi-infinite tubes, where $\rho
\rightarrow \pm\infty$ (the sign coincides with the sign of the
associated residue as we explained in section 2.3). At the ``tip''
of any such tube, the metric (\r{met}) behaves
like the Kasner solution. The number of black holes on each
separate symmetry axis $\rho=0$ is equal to the cyclic period of $d\xi$
along this contour.
\end{enumerate}

In both cases, the metric (\r{met}) near any zero of $d\xi$ describes
a cosmic string just as for $\E =1$; the excess angles are
given by $2\pi k$, where $k$ is the order of the zero, and the
various functions appearing in (\r{met}) are to be evaluated
at this zero.

An artist's view of the section $\phi=const$, $t=const$,
$\rho \geq 0$ of a typical manifold of the second type
is given in Fig.2. Analogous pictures for $d\xi = dh$ would look
precisely like the usual diagrams representing a Riemann
surface as a covering of the complex plane; the only difference is
that here we get a covering of the upper half-plane only. The
number of sheets coincides with the number of black holes and is
equal to the number of poles of $h(P)$.

\section{Action of the Virasoro-Witt generators on the ``partially
discretized'' moduli space}

We now wish to address the question of extra symmetries for the
new solutions that we have constructed in this paper.
Contrary to the axisymmetric solutions studied in the literature
so far, our solutions depend on extra topological degrees of freedom,
namely the moduli of the (punctured) Riemann surface $\L$.
We will show that there is an analog of the solution generating
symmetries acting on these topological degrees of freedom.
Together with the original solution generating symmetries
not affecting the topology of the world-sheet, they form a group
which contains and extends the Geroch group.

To display the new symmetries, we recall that (as shown for instance
in \c{Kont}) it is possible
to define an action of a Virasoro-Witt algebra on the moduli space
of algebraic curves of genus $g$. For all but finitely many
generators, this action simply corresponds to a
reparametrization of the coordinates, and hence has no physical
significance. The remaining generators, on the other hand,
correspond to genuine deformations of the
conformal structure, and thus act non-trivially on the moduli
space, yielding new solutions that cannot be ``reached'' by
the Geroch group. We note that some action of a
Virasoro-Witt algebra on the solutions of the Ernst equation (or, more
precisely, on the spectral parameter plane) was already exhibited
in \c{M2} (cf. also \c{Nic1}).
However, this action
does not give anything new beyond the action of the
Geroch group itself. Rather, its action merely
``stirs'' the singularities of
$\Psi$ in the $\l$-plane. In contrast, the non-trivial action of
the Virasoro-Witt generators in our case is due to the presence
of topological degrees of freedom on the worldsheet (we will
ignore the torus ($g=1$) as it presents no qualitatively new features).

Consider an arbitrary curve $\L$ of genus $g$ and choose some fixed
point $P_0$ on it with related local parameter $\t$ in a local
neighborhood of $P_0$ (so that $\t(P_0) =0$).
On the boundary $\partial D$ of a
small disc $D$ of radius $R$ around $P_0$, we define the vector fields
$$ t_k = - \t^{k}\f{\partial}{\partial\t}\;,\;\;\;\;k\in {\bf Z} $$
Evidently, they generate a Virasoro-Witt algebra $Vir$. The space
of these vector fields can be represented as a direct sum
$$ V=V_-\oplus V_0\oplus V_+ $$
Here, $V_-$ is defined to consist of all vector fields
admitting a holomorphic extension into the disc, i.e. the interior
of the boundary $\partial D$; similarly, $V_+$ consists of all
vector fields possessing a holomorphic extension to the complement
$\L \setminus D$ of the disc, i.e. the exterior of $\partial D$.
The remaining vector fields, which can be extended holomorphically
neither into $D$ nor into its complement, span the subspace $V_0$.
Now it is known \c{Kont} that always
$$ {\bf dim}_{\C} V_0 =3g-3 $$
and, if $P_0$ is not a Weierstrass point on $\L$, a basis
in $V_0$ is given by
\be
t_k \;\;\; , \;\;\;\;k=-1,...,-3g-3
\la{tj}\ee

We are now ready to define the action of an arbitrary vector field
$\v\in V_0$ on the moduli space by the following procedure
\c{GO, Kont}. Cut the curve $\L$ along $\partial D$; this gives two
disconnected parts $D$ and its complement $\L \setminus D$,
whose boundary
points we label by $P_+$ and $P_-$, respectively. We then shift the
boundary points relative to one another after the replacement of
$P_+$ by $\exp ({\beta \v}) P_+$ (i.e. shifting $P_+$ by the
$S^1$ diffeomorphism generated by the vector field $\v$), and
then glue the pieces back together. In general, the conformal
structure of the new curve $\L_{\beta}$ will be different.
According to \c{SS}, the variation of the
$b$-period matrix ${\bf B}$ of $\L$ with respect to this
transformation is described by the formula
$$\f{\partial{\bf B}_{mn}}{\partial\beta}|_{\beta=0} =
\oint_{\partial D}  dU_n dU_m \v $$
where $dU_n$ (for $n=1,...,g$) is a normalized basis of holomorphic
differentials on $\L$ (the expression $ dU_n dU_m\v $ is a
one-form).

Now we see that this variation is zero for $\v\in V_-$
and $\v\in V_+$ because the integration contour $\partial D$
can be deformed to a point if $\v$ admits a holomorphic extension
into $D$, or ``pulled off the back of $\L$'', if there is a
holomorphic extension to the complement of $D$.
Since the matrix of $b$-periods completely determines the
conformal structure of $\L$, a deformation of the conformal
structure requires $\v\in V_0$. Since furthermore
${\bf dim}_{\C} V_0 =3g-3$ precisely coincides with
dimension of the moduli space of $\L$, we conclude that all
generators of $V_0$ vary the conformal structure (observe that
$V_0=V\ominus V_- \ominus V_+ $ is not a subalgebra of $V$, in
contrast to $V_-$ and $V_+$).

For the sake of clarity, let us first consider case where
$d\xi(P)=dh(P)$ for some meromorphic function $h(P)$ on $\L$.
We have to vary the set $(\L,\;h(P))$, i.e. the moduli of
the Riemann surfaces of a given genus $g$ with
$n$ punctures $R_1,...,R_n$. In order to also vary the punctures,
we must add $n$ further generators from $V_+$ to $V_0$,
which must be such that the matrix $s_j(R_i)$ is non-degenerate
(although the value of the determinant of this matrix depends
on the choice of local coordinates at the points $R_j$, the
vanishing or non-vanishing of it is a coordinate-independent
property)
where the new generators are designated by $s_j$ ($j=1,...,n$),
and the related subspace of $V_+$ by $V_n \subset V_+$.
As a result, we can vary the set $(\L,\;R_j)$ by an arbitrary
generator from
$$ V_0\oplus V_n= V\ominus V_-\ominus (V_+\ominus V_n)$$
Of course, this linear subpace of $V$ of dimension $3g-3+n$ is
not a subalgebra, in contrast to $V_-$ and $V_+$. Generically,
we can choose $s_j= t_j,\;\;j= -3g-4,...,-3g-3-n$; then
$V_+\ominus V_n$ is also a subalgebra and $V_0\oplus V_n$ is a
``two-sided'' coset space.

Next consider the more complicated situation, where
$d\xi(P)$ is a differential of the
third kind on $\L$ subject to the discretization conditions
(\r{q1}), (\r{q3}), and we have to vary the set of data
$(\L, d\xi )$. Denote the poles of $d\xi$ by $Q_1,...,Q_n$
and keep the related residues (discretized according to (\r{q1}))
fixed. In analogy to the previous case, we have to add
$n$ generators $s_j\in V_+$ to $V_0$
in order to vary the points $Q_j$,
such that the matrix $s_j(Q_i)$ is non-degenerate. Now, however, to
preserve condition (\r{q3}), we have to keep the periods $A_j\;,\;B_j$
$j=1,...,n$ fixed, since it is impossible to vary the integers $m_j$
and $n_j$ by an infinitesimal transformation. Assuming $A_j$ and $B_j$
fixed, consider the orthogonal decomposition
$$ V_0\oplus V_n =\tilde{V}\oplus\tilde{V}^{\bot} $$
where the generators from $\tilde{V}$ do not vary $A_j$ and $B_j$.
Obviously, since $A_j,B_j\in\R$, we have ${\bf dim}_{\C}
\tilde{V}^{\bot} = g$ and
$${\bf dim}_{\C}\tilde{V} =3g-3+n-g = 2g+n-3 $$
So the linear subspace
$$\tilde{V}=V\ominus V_- \ominus (V_+\ominus V_n)
\ominus\tilde{V}^{\bot} $$
of the full Virasoro-Witt algebra
corresponds to variations of the topological degrees of
freedom of the worldsheet $\L$ in the second case.

The resulting symmetry algebra involving both propagating and
topological degrees of freedom is  the product of
$V_0\oplus V_n$ and the Geroch algebra (i.e. $A_1^{(1)}$) in
the first case, and the product of
the analog of the Geroch algebra and the linear
subspace $\tilde{V}$ in the second case.

\section{Concluding remarks}

There are several open questions that we have barely touched
on and which merit further investigation. Amongst other things,
it is necessary to clarify the structure of the space of
periodic solutions of the Ernst equation and to investigate the
global properties of the Geroch group acting on periodic
solutions. This would permit the verification of our conjecture
that there is a one-to-one correspondence between the asymptotically
flat solutions of the Ernst equation and its asymptotically-Kasner
periodic solutions, and perhaps to answer the
question of whether there arise new and
unphysical singularities in such solutions. Secondly, it would be
desirable to understand in more detail the structure of the
``discretized'' moduli space and the action of the Virasoro-Witt
generators on it.

Our use of string-inspired technology also suggests possible
applications of our results in the context of string theory.
A ``stringy'' interpretation of the metric (\r{ds}) was already
proposed in \c{Nic1}, where it was pointed out that the
conformal factor (Liouville degree of freedom)
$\sigma \equiv k -\f{1}{2}\log(\E+\Eb)$ and the ``dilaton'' field
$\rho$ together appear in the equations of motion in such a way
that they can be interpreted as longitudinal target space degrees of
freedom. All degrees of freedom can be
combined into a Lorentzian ``target space'' metric
\be
ds^2 = d\rho   d\sigma +\rho\frac{d\E d\Eb}{(\E+\Eb)^2}
\la{target}
\ee
such that the equations of motion (\r{E}) and (\r{boxrho})
can be rederived from (\r{target}) by varying with respect to
$\E$ and $\sigma$; the first order equation in (\r{co}) for the
conformal factor and its complex conjugate
are obtained as the ``Virasoro conditions''. Accordingly, the
``matter'' degrees of freedom residing in $\E$ (alias the
transverse polarization states of the graviton) are reinterpreted
as transverse target space degrees of freedom.

Of course, (\r{target}) does not correspond to a conformally
invariant theory because of its explicit dependence on the conformal
factor $\sigma$. However, in Liouville theory as well, the conformal
factor does not decouple, but conformal invariance is nevertheless
restored at the quantum level as the Liouville degree of freedom
``adjusts'' its contribution to the conformal anomaly in the required
manner (see e.g. \c{dhoker} for reviews). Although the viability of
this interpretation of (\r{target}) remains to be tested, we feel
encouraged by the result that the
sigma model based on (\r{target}) is one-loop
finite even though the target space is not Ricci-flat \c{Nic2}
(see also \c{Tseytlin}).

A similar proposal to apply the Ernst equation to the description
of four-dimensional string backgrounds was recently made in
\c{Bakas}, where the target space is taken to be an
axisymmetric stationary four-manifold. The string zero-modes
propagating on this background are an antisymmetric
tensor $B_{\mu \nu}$ and a dilaton (not to be confused with $\rho$
above), whose interactions are governed
(after dualization of $B_{\mu \nu}$) by an $\slr$ sigma model.
In the axisymmetric reduction, the Geroch group is replaced by
the affine Kac Moody extension of
$O(2,2) \cong \slr \times \slr$ in the usual way; it can be interpreted
as a group acting on the space of string backgrounds \c{Bakas}.
The results obtained here suggest that the
existence of even bigger symmetry groups
on the space of string backgrounds which also act on the moduli
of the worldsheet and possibly mix them with the propagating
degrees of freedom.

\bigskip
{\bf Acknowledgments:} D.K. thanks Yavus Nutku for important
discussions.

\newpage
\vspace*{18.0cm}
\begin{center}
{\bf Fig.1:} Integration contour used in the proof of (\r{kp}).
\end{center}
\newpage
\vspace*{18.0cm}
{\bf Fig.2:} Example of a static
solution with two symmetry axes ($\rho=0$)
and three asymptotic regions ($\rho =\i$) for non-vanishing cyclic
periods of $d\xi$. The genus of $\L$ depends on how the surface is
extended to negative $\rho$; for a reflection symmetric
surface, it would be three. The solid points represent cosmic
strings of radii $\rho = \rho_1, \rho_2,...$, and the solid
segments on the contours $\rho =0$ correspond to the horizons
of the respective black holes (not drawn to scale).

\end{document}